\documentclass[12pt]{article}
\usepackage[T1]{fontenc}
\usepackage[all]{xy}
\usepackage{float}
\usepackage{amsmath}
\usepackage{graphicx}

\title{Cavitation controls droplet sizes in elastic media}
\author{Estefania Vidal-Henriquez$^1$ and David Zwicker$^1$}%
\date{%
	$^1$~Max-Planck Institute for Dynamics and Self-Organization, Am Faßberg 17, 37077 Göttingen, Germany.
}
\newcommand{\mean}[1]{\langle #1 \rangle}
\newcommand{\driving}{g}

\renewcommand{\eqref}[1]{[\ref{#1}]}
\newcommand{\Eqref}[1]{Eq.~\eqref{#1}}
\newcommand{\Eqsref}[1]{Eqs.~\eqref{#1}}
\newcommand{\figref}[1]{Fig.~\ref{#1}}
\newcommand{\Rcav}{R_\mathrm{cav}}
\newcommand{\Pcav}{P_\mathrm{cav}}
\newcommand{\PcavMin}{\Pcav^\mathrm{min}}
\newcommand{\Pinf}{P_\infty}
\newcommand{\tCav}{t_\mathrm{cav}}
\newcommand{\cCav}{c_\mathrm{cav}}
\newcommand{\cSat}{c_\mathrm{sat}}
\newcommand{\cEq}{c_\mathrm{eq}}
\newcommand{\cIn}{c_\mathrm{in}}
\newcommand{\tFinal}{t_\mathrm{final}}
\newcommand{\kBT}{k_\mathrm{B} T}

\usepackage{textcomp}
\usepackage[style=phys,backend=bibtex, eprint=true]{biblatex}
\usepackage{units}
\begin{document}
		
	\maketitle
\begin{abstract}
	Biological cells use droplets to separate components and spatially control their interior. Experiments demonstrate that the complex, crowded cellular environment affects the droplet arrangement and their sizes. To understand this behavior, we here construct a theoretical description of droplets growing in an elastic matrix, which is motivated by experiments in synthetic systems where monodisperse emulsions form during a temperature decrease. We show that large droplets only form when they break the surrounding matrix in a cavitation event. The energy barrier associated with cavitation stabilizes small droplets on the order of the mesh size and diminishes the stochastic effects of nucleation. Consequently, the cavitated droplets have similar sizes and highly correlated positions. In particular, we predict the density of cavitated droplets, which increases with faster cooling, as in the experiments.
	Our model also suggests how adjusting the cooling protocol and the density of nucleation sites affects the droplet size distribution. 
	In summary, our theory explains how elastic matrices affect droplets in the synthetic system and it provides a framework for understanding the biological case.
	
\end{abstract}

	Phase separation has emerged as a powerful concept to explain how biological cells structure their interior~\cite{Banani2017,Berry2018}.
	It explains how membrane-less compartments with distinct chemical composition, called biomolecular condensates, form spontaneously.
	In contrast to classical liquid-liquid phase separation, these condensates exist in complex, crowded environments, e.g., provided by the cytoskeleton in the cytosol or the chromatin in the nucleus.
	This fundamentally affects the behavior of condensates:
	their coarsening is slowed down by sub-diffusive motion~\cite{lee2020chromatin}, they are supported against gravity by the F-actin network in the nuclei of large cells~\cite{feric2013nuclear}, and their assembly depends on the stiffness of their surrounding \cite{schwayer2019mechanosensation, kinoshita2020force}.
	Another example are artificially induced condensates, which typically appear in soft regions of the chromatin~\cite{shin2018liquid}.
	Taken together, these experiments and recent numerical simulations~\cite{zhang2020mechanical} demonstrate that condensates react to the elastic properties of their surrounding~\cite{wiegand2020drops}, but the detailed dynamics are still unclear.

	The interaction of droplets with soft elastic matrices can be studied in detail in a synthetic system, where oil droplets are induced in a PDMS matrix by lowering the temperature~\cite{style2018liquid}.
	Similar to the biological case, droplets are biased towards softer regions in this system~\cite{rosowski2020elastic, rosowski2020softmatter}.
	This elastic ripening is absent when the elastic properties of the system are homogeneous.
	Instead, all observable droplets attain similar sizes and their positions are correlated~\cite{style2018liquid}.
	Interestingly, one observes smaller droplets in stiffer systems and at larger cooling rates~\cite{style2018liquid}.
	This implies that the final state is governed by non-equilibrium processes, which is also demonstrated by the bidisperse emulsions that form after increasing the cooling rate during the experiment~\cite{rosowski2020elastic}.

	Theoretical descriptions of such systems have to describe how the elastic matrix affects the droplets' dynamics.
	In the simplest case, the matrix exerts a pressure onto the droplets proportionally to the local stiffness, which is sufficient to explain elastic ripening~\cite{vidal2020theory}.
	Moreover, assuming a strain-stiffening surrounding can explain why droplets attain the same size, which decreases with stiffness~\cite{kothari2020effect, wei2020modeling,ronceray2021liquid}.
	However, these equilibrium models cannot describe the dependence on the cooling rate.
	
	In this paper, we present a dynamic theory of droplet formation in elastic matrices, which is based on the assumption that droplets can break the surrounding matrix.
	We show that in this case some droplets cavitate and grow macroscopically, while a large fraction is restricted to mesh size.
	The cavitated droplets have a similar size, which decreases with larger cooling rate.
	We motivate our theory by first considering how the elastic matrix affects a single droplet.
	We then couple the dynamics of multiple droplets via the diffusion of monomers in the dilute phase.
	Using numerical simulations and analytical approximations, we demonstrate that this model can explain all the experimental observations of the synthetic system.

	\section*{External pressure governs dynamics of droplets}
	To understand how droplets interact with an elastic matrix, we first consider the free energy of a single spherical droplet of radius~$R$. 
	Droplet growth is driven by the differences in chemical potential and osmotic pressure between the droplet and its surrounding.
	We show in the SI that this can be captured by a driving strength~$\driving$, which quantifies the energy gain when the droplet volume~$V=(4\pi/3) R^3$ increases.
	However, when the droplet grows its surface area~$A=4\pi R^2$ also increases, which comes at a cost proportional to the surface energy~$\gamma$.
	
	Moreover, the matrix surrounding the droplet must be displaced, which we capture by an elastic energy $F_E(V)$.
	Taken together, the free energy~$F$ of the entire system reads 
	\begin{equation}
	F = -\driving\, V + \gamma A + F_E(V) \; ,
	\label{Eq:Free_energy_noP}
	\end{equation}
	where we for simplicity first consider constant driving strength~$\driving$ and surface energy~$\gamma$.

	\begin{figure}
		\centering
		
		\includegraphics{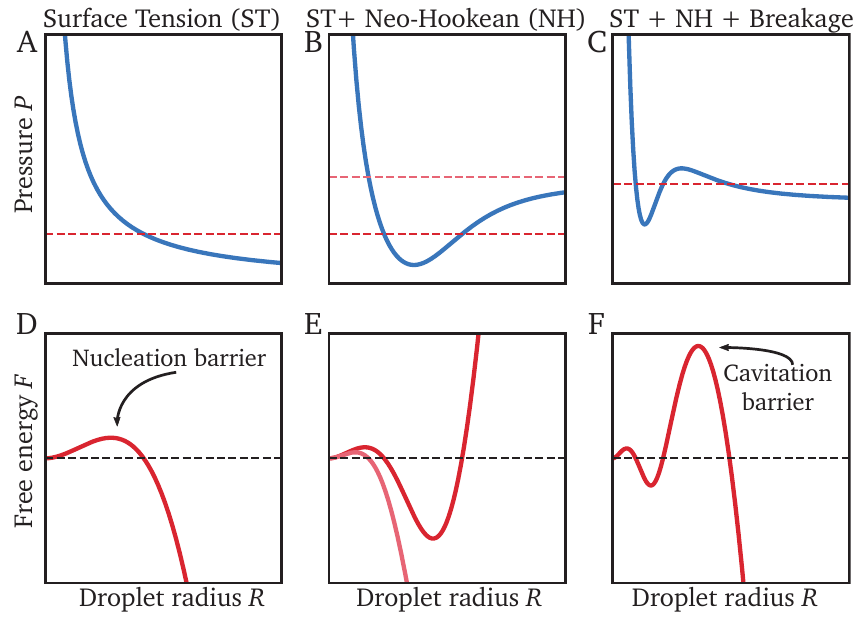}
		
		\caption{\textbf{Breakage implies a cavitation barrier}
			Example of pressure $P= P_E +P
			_\gamma$ exerted on a droplet during growth (upper panels) and corresponding free energy~$F$ (lower panels) as a function of the droplet radius~$R$ for three different scenarios:
			Without an elastic mesh ($P_E = 0$), the pressure curve is monotonously decreasing (Panel A). After crossing a nucleation barrier, droplets grow until all material is absorbed (Panel D).
			An elastic matrix increases the pressure once the droplet grows beyond mesh size (Panel B).
			This can lead to an energy minimum where droplets are stable (Panel E).
			If the mesh can break, the pressure curve exhibits a local maximum (Panel C), which leads to a cavitation barrier (Panel F).
		}
		\label{Fig:Stress_strain}
	\end{figure}
	
	A droplet will grow spontaneously when the free energy decreases ($\partial F/\partial V<0$), i.e., if 
	\begin{equation}
	\driving > P(R)
	\qquad \text{with} \qquad
	P(R) =P_\gamma(R) + P_E(R)\;,
	\label{Eq:Growth_condition}
	\end{equation}
	where $P_\gamma = 2\gamma/R$ is the Laplace pressure due to the surface tension~$\gamma$~\cite{Gauss1877, neumann1894} and $P_E=\partial F_E/\partial V$ is the pressure~ exerted by the elastic matrix; see SI.
	A droplet thus grows when the driving strength $\driving$ exceeds the pressure~$P$ exerted on the droplet.
	A stationary state with droplet radius $R^*$ is reached when $\driving = P(R^*)$, which is stable if $\partial^2F/\partial V^2>0$, or,
	\begin{equation}
	P'(R_*)>0
	\;.
	\label{Eq:Stability_criterion}
	\end{equation}
	A droplet is thus stable when the exerted pressure increases with its size.
	
	Without an elastic matrix, the droplet is only affected by the Laplace pressure $P_\gamma$; see \figref{Fig:Stress_strain}A.
	The corresponding free energy shown in \figref{Fig:Stress_strain}D demonstrates that surface tension dominates for small droplets.
	In particular, droplets can only grow spontaneously ($\partial F/\partial V<0$) after overcoming a \emph{nucleation barrier}, e.g., by thermal fluctuations (homogeneous nucleation)\cite{turnbull1950kinetics} or thanks to nucleation sites that lower the barrier (heterogeneous nucleation); see SI.
	Once the droplet is big enough, the energy decreases with increasing radius and the droplet is always unstable ($P_\gamma'(R) <0$).
	Droplet growth is then only restricted by the available amount of material.
	
	\section*{An elastic matrix restricts droplet growth}
	
	An elastic matrix surrounding the droplet exerts an additional pressure and thus potentially opposes growth; see \Eqref{Eq:Growth_condition}. 
	The pressure exerted by the matrix depends on its elastic response. 
	For small deformations, the response can be characterized by the Young's modulus~$E$.
	However, droplets can grow much larger than the mesh size $\ell$, implying large deformations of the matrix.
	The simplest model describing such hyperelastic material is the Neo-Hookean model, where the pressure on a spherical cavity of radius $R$ is monotonically increasing ($ P_E'(R)>0$) and converges at large radii to $P_E=5E/6$ \cite{mooney1940theory}.
	If the driving strength $\driving$ is lower than the maximal pressure, the system opposes further droplet growth and leads to a stable radius~$R_*$ when $\driving = P(R_*)$; see \figref{Fig:Stress_strain}B.
	This steady state corresponds to a minimum in the free energy; see \figref{Fig:Stress_strain}E.
	Therefore, an elastic mesh providing resistance to droplet growth can stabilize droplets.
	
	\section*{Breakage provides a cavitation barrier for droplets}
	The Neo-Hookean model is often too simple to describe realistic materials, in part because it does not account for breaking bonds in the elastic mesh.
	To capture breakage, we next consider a stress-strain curve that has a maximal pressure $P_\mathrm{cav}$ at a finite radius $R_\mathrm{cav}$; see \figref{Fig:Stress_strain}C.
	Similar to the Neo-Hookean model, we consider an increasing pressure when droplets grow beyond the mesh size~$\ell$.
	However, at the critical radius~$R_\mathrm{cav}$ the mesh cannot sustain the stress anymore and breaks, resulting in a pressure decrease~\cite{raayai2019intimate}.
	The stability criterion given in \Eqref{Eq:Stability_criterion} indicates that droplets with $R=R_\mathrm{cav}$ are unstable and will thus expand rapidly in a cavitation event~\cite{barney2020cavitation,gent1969nucleation}.
	
	The non-monotonous stress-strain relation results in a free energy that has two energy barriers; see \figref{Fig:Stress_strain}F.
	The first barrier is the familiar nucleation barrier, while the second one is the \emph{cavitation barrier}.
	The local minimum between the two barriers corresponds to the stable state described in the case of the Neo-Hookean model.
	However, with breakage, droplets can overcome the second barrier and cavitate if the driving strength $\driving$ exceeds $P_\mathrm{cav}$.
	The growth of such droplets would then only be limited by the available amount of material, similar to the case without any elastic matrix.

	\section*{Multiple droplets grow when temperature is decreased}

	In the experiments of Style et al.~\cite{style2018liquid}, multiple oil droplets appeared simultaneously when the temperature was lowered.
	Since lowering the temperature corresponds to increasing the driving strength~$\driving$, droplets appear when $\driving$ reaches the maximal pressure exerted by the surrounding matrix.
	If this maximal pressure increases with the overall stiffness, we predict that lower temperatures are necessary to create droplets in stiffer systems, which was indeed observed~\cite{rosowski2020elastic}.
	However, this qualitative analysis does not distinguish between the Neo-Hookean and the Breakage model, since both provide a maximal pressure that explains the simultaneous growth of droplets. 
	
	To distinguish the Neo-Hookean from the Breakage model, we need to analyze the droplet dynamics in detail.
	Since the elasto-adhesive length scale of the experimental system is smaller than the droplet size~\cite{kim2020extreme}, elastic interactions of droplets are negligible.
	In contrast, growing droplets compete for the material dissolved in the dilute phase, which couples their dynamics.
	We analyze this using a mean-field theory, where we describe a collection of immobile droplets by their positions $\vec{x}_i$ and their radii~$R_i$ together with the concentration field $c(\vec x)$ in the dilute field~\cite{vidal2020theory}.
	For simplicity, we assume that the concentration $\cIn$ inside each droplet is constant and that droplets are in equilibrium with their immediate surrounding, which exerts the pressure~$P(R)$ onto the droplet.
	This implies that the concentration right outside the interface of a droplet is given by \cite{vidal2020theory}
	\begin{equation}
	\cEq(P,T)=\cSat(T)
	\exp\left({\dfrac{P}{\cIn  k_\mathrm{B}T}}\right)
	\;,
	\label{Eq:Phi_eq}
	\end{equation}
	where $\cSat$ is the equilibrium concentration in the absence of an elastic mesh in the thermodynamic limit, $k_B$ is Boltzmann's constant, and $T$ is the system's temperature. 
	
	Droplets grow when their surrounding is supersaturated ($c > \cEq$).
	The droplet growth rate reads \cite{vidal2020theory}
	\begin{equation}
	\frac{\mathrm{d}R_i}{\mathrm{d}t} =
	\frac{D}{R_i \cIn}\left[
	c(\vec{x}_i)-\cEq\bigl(P(R_i), T\bigr)
	\right]
	\;,
	\label{Eq: radial_dynamics}
	\end{equation}
	where $D$ is the diffusivity of the droplet material in the dilute phase.
	The concentration in the dilute phase obeys
	\begin{equation}
	\partial_t c = D \nabla^2 c - \cIn\sum_i \dfrac{\mathrm{d}V_i}{\mathrm{d}t} \delta(\vec{x}_i - \vec{x})
	\;,
	\label{Eq: dilute_field_dynamics}
	\end{equation}
	where the last term accounts for material exchange with the droplets~ \cite{vidal2020theory}.
	Taken together with no-flux conditions at the system's boundary, Eqs.~[\ref{Eq: radial_dynamics}--\ref{Eq: dilute_field_dynamics}] conserve the total amount of droplet material.
	
	We simulate the system by mimicking the experimental protocol of Style et al.~\cite{style2018liquid}.
	In particular, we consider a linear relation between the saturation concentration~$c_\mathrm{sat}$ and temperature~$T$ together with a constant cooling rate.
	Consequently, $c_\mathrm{sat}$ decreases linearly from the initial value $c_0$ until it reaches the minimal value $c_0 - \Delta c$ at the final temperature,
	\begin{equation}
	c_\mathrm{sat}(t)= 
	\begin{cases}
	c_0 - \alpha t & t < \frac{\Delta c}{\alpha} \\
	c_0 - \Delta c & \text{otherwise}    \;,
	\end{cases}
	\label{Eq:quench}
	\end{equation}
	where $\alpha$ is the rate of the decreases.
	As $\cSat$ is lowered, the equilibrium concentration~$\cEq$ also decrease, see \Eqref{Eq:Phi_eq}, implying a larger supersaturation $c-\cEq$, which corresponds to a higher driving strength~$\driving$.
	Starting with a homogeneous system at high temperature (high $\cSat$), droplets will nucleate once $\driving$ is large enough to cross the nucleation barrier; see \figref{Fig:Stress_strain}D--F.
	In the experimental system, homogeneous nucleation is basically impossible and droplets must thus nucleate heterogeneously at nucleation sites; see SI.
	This suggests that surface tension effects are negligible for small droplets.
	In fact, surface tension is also negligible for large droplets since Ostwald ripening is slow  (see SI), suggesting that the total pressure $P$ is always dominated by the elastic pressure~$P_E$.
	We thus neglect surface tension for simplicity and rather assume that droplets form quickly at nucleation sites.
	In particular, we initialize our simulations with many small droplets with radii on the order of the mesh size, $R_i(t=0) = \ell$, and focus on the subsequent dynamics.
	The radius at which droplets are initialized is unimportant, since they are restricted by the elastic matrix to have a small radius~$R$ that is governed by the condition $g=P(R)$.
	Consequently, many microscopic droplets coexist early in the simulation.
	
	Droplets can grow macroscopically ($R \gg \ell$) when the driving strength~$\driving$ exceeds the pressure exerted by the mesh.
	Since realistic meshes are heterogeneous~\cite{malo2015heterogeneity}, the exerted pressure will vary slightly from droplet to droplet.
	To capture such heterogeneity for the Neo-Hookean model (NH), we consider variable mesh sizes~$\ell_i$,
	\begin{equation}
	P^\mathrm{NH}_i(R) = E\left(
	\frac56 - \frac{2\ell_i}{3R} - \frac{\ell_i^4}{6R^4}
	\right)
	\label{Eq:NeoHookean_model}
	\;,
	\end{equation}
	where $E$ is the macroscopic Young's modulus of the material.
	Conversely, in the Breakage model (BR), we choose random cavitation pressures~$\Pcav^{(i)}$, since this parameter dominates the cavitation barrier.
	We thus consider the simple form
	
	\begin{equation}
	P^\mathrm{BR}_i(R)=
	\begin{cases}
	0 & R<\ell\\
	\Pcav^{(i)}\dfrac{R-\ell}{\Rcav - \ell} 
	&\ell\leq R\leq \Rcav\\
	\Pinf & R>\Rcav
	\;,
	\end{cases}
	\label{Eq:Breakage_model}
	\end{equation}
	where we keep both $\ell$ and $\Rcav$ fixed for all droplets, since varying these parameters does not affect the results significantly; see SI.
	\Eqref{Eq:Breakage_model} implies that the external pressure increases linearly when the droplet grows beyond the mesh size~$\ell$ until it reaches the cavitation radius~$R_\mathrm{cav}$.
	Beyond this threshold, the mesh breaks and provides a constant resistance quantified by a pressure $P_\infty < \Pcav^{(i)}$.

	\begin{figure}[tb]
		\centering
		
		\includegraphics{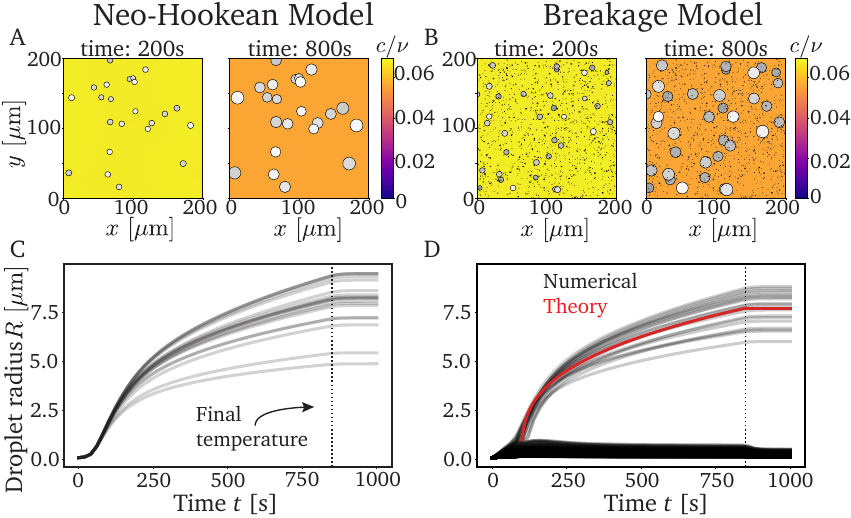}
		\caption{\textbf{Decreasing temperatures cause monodisperse emulsions} in the Neo-Hookean model (NH, Panels A,C) and Breakage model (BR, Panels B,D).
			(A-B) 2d projections of typical simulations at two time points.
			The heat map indicates the concentration~$c$ in the dilute phase and droplets are marked with disks where color saturation indicates depth.
			(C-D) Droplet radii $R$ as a function of time~$t$ showing that large droplets are monodisperse after reaching the final temperature at $t=\Delta c/\alpha$ (dotted black line). 
			The red dashed line shows the theoretical prediction given by \Eqref{Eq:Droplet_long_term_growth}.
			The model parameters are
			$\alpha$ = \unit[1.864 $\cdot$ 10\textsuperscript{-5}]{s\textsuperscript{-1}$\nu$\textsuperscript{-1}}, $E=$\unit[186]{kPa},
			$\Delta c =$ 0.0159 $\nu^{-1}$, 
			$D$ = \unit[50]{\textmu m\textsuperscript{2} s\textsuperscript{-1}}, $c_\textrm{in} = \nu^{-1}$, and
			$c_\mathrm{in} \kBT=$\unit[11]{MPa}.
			For the NH model, we sample $\ell$ uniformly between \unit[0.1]{\textmu m } and \unit[0.102]{\textmu m }~\cite{malo2015heterogeneity} and use $m$ = \unit[7 $\cdot$ 10\textsuperscript{-6}]{\textmu m\textsuperscript{-3}}.
			For the BR model, we have $R_\mathrm{cav} =$ \unit[1]{\textmu m },  $m$ = \unit[1.875 $\cdot$ 10\textsuperscript{-4}]{\textmu m\textsuperscript{-3}}, $\eta/m =$ \unit[10\textsuperscript{5}$E$]{ \textmu m\textsuperscript{3}}, $P_\mathrm{cav}^\mathrm{min}=$ $E$, and $P_\infty=$ 5/6$E$.
		}
		\label{Fig:NH_vs_Breakage}
	\end{figure}

	\figref{Fig:NH_vs_Breakage} shows typical simulations of \Eqsref{Eq:Phi_eq}--\eqref{Eq:quench} for both the Neo-Hookean model (\Eqref{Eq:NeoHookean_model}) and the Breakage model (\Eqref{Eq:Breakage_model}).
	In both cases, macroscopic droplets appear and they grow with very similar rates.
	However, the Breakage model additionally exhibits a large number of microscopic droplets, which apparently do not grow.
	Since these microscopic droplets are likely not visible in the experiment, both models appear to yield mono-disperse emulsions, although this requires an extremely homogeneous mesh in the Neo-Hookean model.
	In contrast, the models behave differently when we nucleate new droplets during the simulation: While all newly nucleated droplets grow in the Neo-Hookean model, in the Breakage model most droplets are restricted to microscopic sizes; see SI.
	Consequently, we expect that the Breakage model leads to a more uniform size distribution of large droplets in realistic situations.
	
\begin{figure}[h!]
		\centering
		\includegraphics[width=\textwidth]{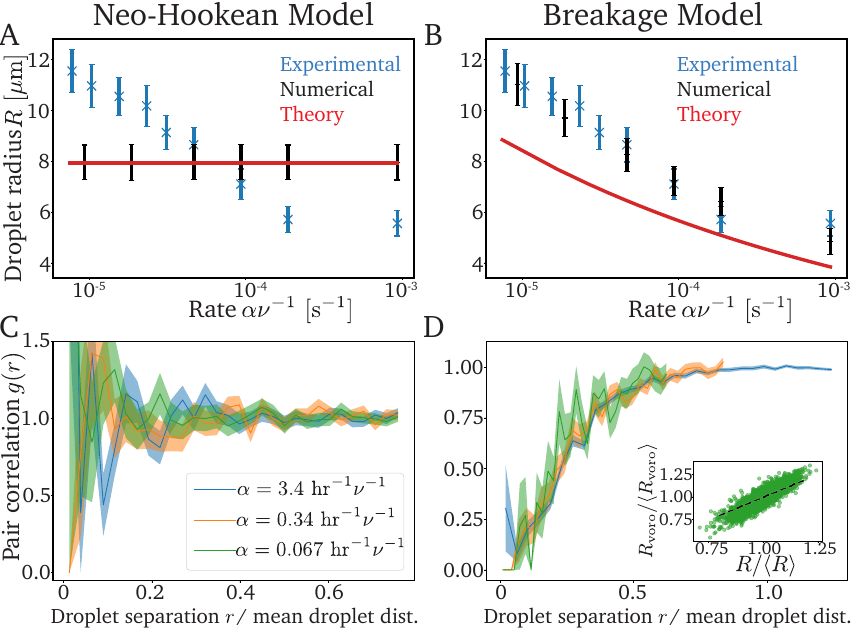}
		\caption{\textbf{The Breakage model explains the experimental data.}
			(A,B) Comparison of the experimental data (blue dots, \cite{style2018liquid}), numerical simulations (black symbols), and analytical predictions (red lines) for the radius~$R$ of the cavitated droplets as a function of the rate~$\alpha$ with which the saturation concentration decreases for the Neo-Hookean model and the Breakage model.
			(C,D)
			Scaled pairwise correlation function $g(r)$ of the cavitated droplets (see SI) for three rates~$\alpha$.
			The inset in panel D shows the correlation between the radius~$R$ of droplets and the size $R_\mathrm{voro}$ of their surrounding, which is obtained from a Voronoi tesselation.
			(A-D) Model parameters are as in Fig.~\ref{Fig:NH_vs_Breakage}, except for $\eta/m=$ \unit[3$\cdot$10\textsuperscript{5}$E$]{\textmu m \textsuperscript{3}}.
		}
		\label{Fig:Exp_comparison}
	\end{figure}

	To see which of the two models provide a better explanation of the experiments, we next test their predictions quantitatively.
	Here, we use the experimentally measured values of $D$, $\Delta c$, $\alpha$, $P_\infty$, $E$, and $c_\textrm{in}$, while the values of the mesh size~$\ell$ and the cavitation radius~$\Rcav$ are arbitrary and do not affect the predictions of the model; see SI.
	The only relevant parameter, which we adjust to match the experimental data, is $\eta/m$ quantifying the mesh heterogeneity and the density of nucleated droplets.
	We first focus on the intriguing non-equilibrium effect that larger cooling rates lead to more and smaller droplets.
	In the Neo-Hookean model, the average droplet size~$\mean{R}$ is independent of the cooling rate~$\alpha$ (see \figref{Fig:Exp_comparison}A), while it matches the experimental data in the Breakage model (\figref{Fig:Exp_comparison}B), including the dispersion statistics; see SI.
	The two models also differ in the spatial distribution of large droplets, which we quantify by the pair correlation function, similar to the experiments~\cite{style2018liquid}.
	\figref{Fig:Exp_comparison}C shows that droplets are uniformly distributed in the Neo-Hookean model since their positions are solely controlled by their nucleation.
	In contrast, droplet cavitation seems to be correlated in the Breakage model (\figref{Fig:Exp_comparison}D), leading to a low probability of finding two large droplets close to each other, similar to the experiments~\cite{style2018liquid}.
	The shown data collapse suggest that the pair correlation function is scale-free.
	Moreover, the volume surrounding a droplet, measured from a Voronoi tessellation, is strongly correlated with its size; see inset of \figref{Fig:Exp_comparison}D.
	The fact that our simulations match the experimental data quantitatively suggests that breakage is a crucial aspect.

	\section*{Large droplets suppress further cavitation by depleting their vicinity}
	To understand how breakage affects the droplets' dynamics, we next investigate why some droplets cavitate while others remain small; see \figref{Fig:NH_vs_Breakage}D.
	Initially, all droplets are small and grow due to the decreasing saturation concentration by absorbing the excess material from the dilute phase.
	Note that droplets in a softer environment, i.e., with a lower $\Pcav$, exhibit a lower equilibrium concentration~$\cEq$, see \Eqref{Eq:Phi_eq}, and thus grow faster.
	This initial growth phase continues until the droplet with the lowest $\Pcav$ reaches its cavitation radius~$\Rcav$.
	At this point, the elastic matrix no longer provides enough resistance ($\partial P/\partial R < 0$) and the droplet radius becomes unstable; see \Eqref{Eq:Stability_criterion}.
	The droplet thus cavitates by recruiting material from the dilute phase as fast as possible in a diffusion limited process.
	Such a quickly growing droplet depletes its surrounding, effectively fixing the local driving strength to $g = P_\infty$.
	Consequently, other droplets in the vicinity cannot cavitate and will remain small forever.
	Taken together, the growth of a cavitated droplet prevents the cavitation of other droplets in its surrounding while droplets further away might still grow, which qualitatively explains the observed pair correlation function; see     \figref{Fig:Exp_comparison}D.

	The numerical data shown in \figref{Fig:NH_vs_Breakage}D. suggests that all droplets that become large cavitated at very similar times~$t=\tCav$ and grow with similar rates.
	To understand the growth dynamics, we first consider cavitated droplets that are homogeneously distributed with a number density $n$.
	Assuming that the cavitated droplets absorb all excess material from the dilute phase, we predict their volume to increase as
	\begin{equation}
	V(t) = V_\mathrm{cav} + \dfrac{\alpha}{n \cIn}(t -\tCav)
	\;,
	\label{Eq:Droplet_long_term_growth}
	\end{equation}
	where $V_\mathrm{cav} = (4\pi/3)\Rcav^3$.
	The dashed line in \figref{Fig:NH_vs_Breakage}D shows that the equivalent prediction for the droplet radius explains the mean growth dynamics of cavitated droplets.
	In fact, this analysis is also valid for the Neo-Hookean model shown in \figref{Fig:NH_vs_Breakage}C since droplets also start growing around the same time and absorb all excess material in this case.
	Taken together, this analysis indicates that the large droplets are mono-disperse because they start growing at the same time and grow with the same rate.
	However, while these conditions are met artificially by our setup of the Neo-Hookean model, they are self-organized in the Breakage model by controlling which droplets cavitate.
	
	The final droplet size can be estimated by evaluating \Eqref{Eq:Droplet_long_term_growth} at the time $\tFinal=\Delta c/\alpha$ when the final temperature is reached.
	For simplicity, we consider the case where droplets are large compared to the cavitation threshold~$\Rcav$, which also implies $\tCav \ll \tFinal$ and leads to $V_\mathrm{final} \approx \Delta c/(n\cIn)$.
	This approximation correctly predicts that the final droplet volume is independent of the quench rate~$\alpha$ in the Neo-Hookean model where the droplet density~$n$ is set by the initial condition; see \figref{Fig:Exp_comparison}A.
	Conversely, in the Breakage model, the density of cavitated droplets might depend on the quench rate~$\alpha$, which could explain the observed size-dependence shown in \figref{Fig:Exp_comparison}B.

	\section*{Number and size of cavitated droplets depend on quench rate and cavitation thresholds}
	
	To understand why faster cooling leads to more and smaller droplets, we next focus on the cavitation process in the Breakage model.
	Since cavitated droplets suppress further cavitation in their vicinity, we hypothesize that this suppression is less efficient when the system is cooled faster, implying that more droplets can  cavitate overall.
	
	To estimate the final density~$n$ of cavitated droplets, we analyze a simplified theoretical model.
	The main idea is to study a fixed density~$n$ of cavitated droplets and test whether additional droplets could cavitate in this situation.
	The best estimate is then the lowest value of $n$ where no more droplets cavitate.
	For simplicity, we consider a homogeneous distribution of cavitated droplets, allowing us to focus on a single droplet of radius $R=\Rcav$ in a spherically symmetric domain of volume $n^{-1}$.
	We then obtain the concentration field~$c(r)$ around the droplet by solving the diffusion equation with the boundary condition $c(\Rcav)=\cEq(P_\infty, T)$; see SI.
	Additional cavitation takes place in the dilute phase if there is a droplet whose critical concentration $\cCav=\cEq(\Pcav, T)$ is lower than the actual concentration~$c$ at its position.
	Note that the cavitation pressures~$\Pcav$ are randomly distributed since the elastic matrix is heterogeneous.
	However, since cavitation only happens for low $\Pcav$, it is sufficient to specify the associated cumulative distribution function $\mathcal{F}(\Pcav)$ to linear order around the lower bound $\PcavMin$,
	\begin{equation}
	\mathcal{F}(\Pcav) = 
	\frac{\Pcav - \PcavMin}{\eta}\,\Theta(\Pcav - \PcavMin) \;,
	\label{Eq:Cummulative_distribution}
	\end{equation}
	where $\Pcav - \PcavMin \ll \eta$.
	Here, $\Theta(x)$ is Heaviside's function and $\eta$ describes how widely the small cavitation pressures are distributed. $\eta$ thus quantifies the heterogeneity of the mesh. 
	Considering a homogeneous density~$m$ of nucleated droplets, we can then calculate the expected value of droplets that cavitate in the volume~$n^{-1}$.
	This theory is self-consistent if exactly one droplet cavitates in this volume, which provides an implicit condition for the sought density~$n$ of cavitated droplets; see SI.

	\begin{figure}
		\centering
		\includegraphics[width=\linewidth]{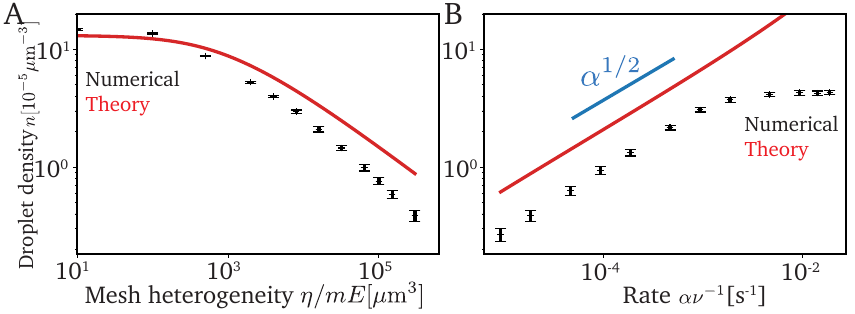}
		
		\caption{\textbf{Suppression of cavitation by large droplets explains numerical data.}
			(A) Density~$n$ of cavitated droplets from numerical simulations (black symbols) compared to the analytical prediction (red line) as a function of the mesh heterogeneity~$\eta$.
			(B) $n$ as a function of the rate $\alpha$ with which the saturation concentration decreases.
			(A,B) Model parameters are given in Fig. \ref{Fig:NH_vs_Breakage}, except $\eta/m$ = \unit[3$\cdot$10\textsuperscript{5}$E$]{\textmu m\textsuperscript{3}} in panel B.
		}
		\label{Fig:Density_vs_eta_and_quench}
	\end{figure}
	
	The theory does not have any adjustable parameters and we thus compare it directly to our numerical simulations.
	\figref{Fig:Density_vs_eta_and_quench}A shows that the density~$n$ of cavitated droplets decreases when fewer droplets nucleate (smaller $m$) or cavitation thresholds~$\Pcav$ are wider distributed (more heterogeneous network, higher $\eta$).
	This is because these two parameters define how many nucleated droplet possess a low enough threshold to cavitate.
	Conversely, \figref{Fig:Density_vs_eta_and_quench}B shows that more droplets cavitate when the system is cooled faster.
	Since the total amount of material taken up by droplets is conserved, this implies smaller droplets for faster cooling, consistent with \figref{Fig:Exp_comparison}B.
	While our theory shows the same trends as the numerical simulations, it consistently overestimates $n$ by roughly a factor of $2$ in most cases.
	This is likely because we assumed a homogeneous distribution of the droplets with the lowest cavitation threshold, while in reality two droplets with low threshold might out-compete each other, effectively leading to a higher cavitation threshold than we anticipate.
	However, our theory indicates that the cavitated droplets deplete the dilute phase, thus suppressing further cavitation. Since this depletion is diffusion-limited, decreasing temperature slowly implies stronger suppression, leading to fewer and larger droplets.

	\section*{Increasing cooling rates cause bidisperse emulsions}
\begin{figure}
		\centering
		\includegraphics[width=\linewidth]{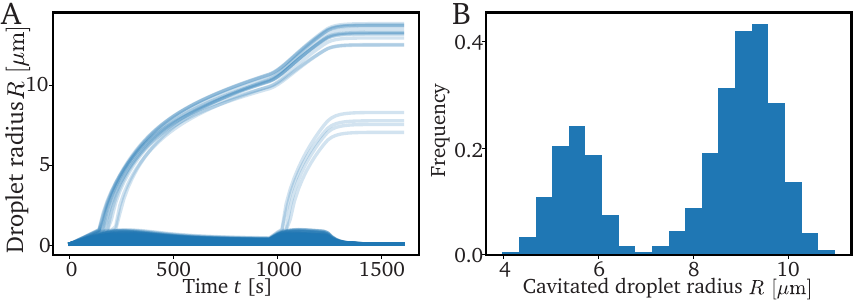}
		\caption{ \textbf{Increasing cooling rate yields bidisperse emulsion.}
			Shown are the droplet radii as a function of time (A) and the droplet size distribution (B) from a numerical simulation with $\eta/m =$ \unit[3$\cdot$10\textsuperscript{5}$E$]{ \textmu m \textsuperscript{3}}, $E =$ \unit[80]{kPa}, and $\alpha =$ \unit[7.77$\cdot$10\textsuperscript{-6}]{s\textsuperscript{-1}$\nu$\textsuperscript{-1}} for $t<\unit[960]{s}$, then $\alpha =$ \unit[3.11$\cdot$10\textsuperscript{-5}]{s\textsuperscript{-1}$\nu$\textsuperscript{-1}}. Other parameters as in Figure \ref{Fig:NH_vs_Breakage}. }
		\label{Bimodal_plots}
	\end{figure}

	We showed that the number and size of the cavitated droplets depends on the depletion of the dilute phase and thus the cooling rate.
	This implies that additional droplets could cavitate when the cooling rate is increased, while lowering the cooling rate should merely slow down droplet growth.
	Indeed, experiments by Rosowski et al. showed a bimodal droplet size distribution when the cooling rate was rapidly increased in the middle of the experiment~\cite{rosowski2020elastic}.
	To explain this observation, we perform a numerical simulation where we rapidly increase the cooling rate well after the first generation of droplets has cavitated.
	This results in a second generation of cavitated droplets, which then grow together with the previously cavitated ones; see \figref{Bimodal_plots}.
	We show in the SI that other size distributions are possible when the rate is changed multiple times.
	Taken together, this demonstrates that different droplet size distributions can be engineered by adjusting the cooling protocol.
	
	\section*{Heterogeneous nucleation might
		explain more cavitated droplets in stiffer systems}
	So far, we have investigated how the density and sizes of the observed droplets depend on the cooling rate~$\alpha$.
	Another important observation of Style et al. \cite{style2018liquid} is that the droplet density~$n$ increases linearly with the Young's modulus~$E$ of the elastic matrix.
	This implies that stiffer matrices lead to smaller droplets.
	Unfortunately, it is difficult to connect $E$, which measures the macroscopic response of the matrix to small strains, to the microscopic details required by our model.
	We thus next consider several possibilities to elucidate which microscopic picture could explain the experimental data.
	
\begin{figure}
		\centering
		\includegraphics[width=\linewidth]{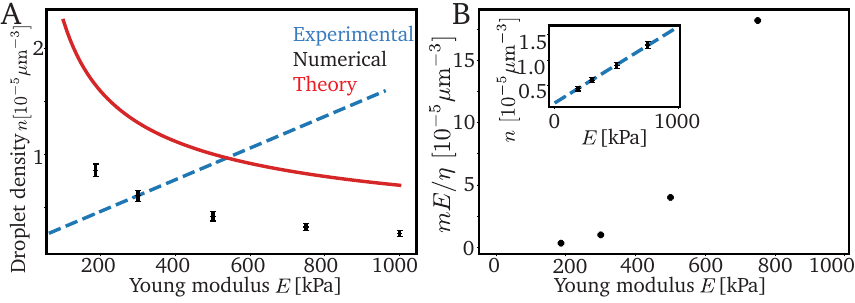}
		\caption{\textbf{Increasing nucleation density could explain stiffness dependence}
			(A)~Droplet density~$n$ as a function of Young's modulus~$E$.
			Our numerical (black symbols) and analytical (red line) theory, based on a linear scaling of pressures with $E$, cannot explain the experimental data (blue dashed line, \cite{style2018liquid}).
			(B) Predicted nucleation site density~$m$ as a function of $E$ to match the measured $n(E)$ shown in the inset. Inset: $n(E)$ from experiments (blue dashed line, \cite{style2018liquid}) and numerical simulations (black symbols).
			(A,B) Model Parameters are $\alpha =$ \unit[2.33$\cdot$10\textsuperscript{-5}]{s\textsuperscript{-1}$\nu$\textsuperscript{-1}}, and given in \figref{Fig:NH_vs_Breakage}}
		\label{Only_stiffness}
	\end{figure}

	In the simplest case, the bulk modulus~$E$ is connected to the pressure curve $P(R)$.
	For example, the Neo-Hookean model implies $P^\mathrm{NH}(R\rightarrow\infty) = \frac56 E$; see \Eqref{Eq:NeoHookean_model}.
	The relation is more complicated for the breakage model, but recent experiments~\cite{raayai2019intimate} indicate that both the cavitation pressure~$\Pcav$ as well as the pressure~$\Pinf$ exhibited by large droplets scale with~$E$.
	Using this scaling in our model, we obtain slightly smaller droplet densities for stiffer matrices, opposite to what we expect from the experiments; see \figref{Only_stiffness}A. 
	Consequently, the scaling of the pressure with~$E$ cannot explain the observed data.
	
	Our model would yield more (and smaller) cavitated droplets when the density~$m$ of nucleated droplets was increased.
	We thus speculate that stiffer systems nucleate more droplets.
	Indeed, we can explain the observed linear increase of the density $n$ of cavitated droplets with $E$ by postulating that $m$ strongly increases with $E$; see \figref{Only_stiffness}B.
	So far, it is not clear how droplets actually nucleate in the elastic network, but it is likely that heterogeneous nucleation plays a role.
	For instance, the cross-linking molecules that are used to create the PDMS matrix could act as nucleation sites.
	In this case, stiffer gels would have more nucleated droplets simply because they contain more cross-linkers~\cite{style2018liquid, rosowski2020elastic, raayai2019intimate}.
	Moreover, stiffer networks might be more homogeneous~\cite{malo2015heterogeneity}, which would be capture by a smaller mesh heterogeneity~$\eta$.
	Taken together, these two effects might explain our prediction that the parameter $m/\eta$ increases strongly with $E$.

	\section*{Conclusions}
	
	We identified a novel mechanism to create monodisperse emulsions, where some growing droplets break the surrounding elastic matrix in a cavitation event. 
	While these droplets become macroscopic, most droplets stay constrained by the matrix and do not grow significantly beyond mesh size.
	The cavitation barrier imposed by the elastic matrix thus separates the stochastic nucleation phase from a deterministic growth phase.
	The resulting cavitated droplets have correlated positions and similar sizes, which can be controlled by the cooling rate.
	Our model agrees quantitatively with experiments~\cite{style2018liquid, rosowski2020elastic} and it suggests how this mechanism can be used to create microscopic patterns in technological applications.

	Monodisperse emulsion also emerge in other situations of driven phase separation.
	For instance, supplying more droplet material externally~\cite{Vollmer2014}, internally using solubility gradients~\cite{Weber2017}, or by chemical reactions~\cite{Zwicker2015} all lead to narrower droplet size distributions than expected from the standard Lifshitz-Slyozov argument~\cite{Lifshitz1961}.
	In all these cases, the diffusive flux between droplets that normally drives Ostwald ripening is dominated by the flux of the supplied droplet material.
	In our system, all droplets additionally start growing at similar times, because they cross the cavitation barrier at similar saturation concentration.
	Taken together, this ensures that droplets reach similar sizes, despite multiple opposing processes:
	Beside the heterogeneities in the elastic properties that causes the dispersion in our model, thermal fluctuations might also contribute.
	Moreover, both Ostwald ripening, driven by surface tension, and Elastic ripening, driven by stiffness gradients over long length scales~\cite{rosowski2020elastic,vidal2020theory}, will affect the droplet size distribution in realistic systems.
	It will be interesting to study all these interactions in the future.

	We expect similar behaviors for biomolecular condensates, which often form as a response to changes in temperature, pH, salt concentration, or protein concentration in cells~\cite{Hyman2014, Alberti2019, Lyon2020,diaz2015temperature}.
	Moreover, chemical modifications, like post-translational modifications, allow cells to actively regulate condensates~\cite{Hondele2020, Soeding2019}.
	All these changes could in principle drive droplet formation, similar to the cooling in our example.
	Biomolecular condensates are also typically constrained by elastic matrices~\cite{wiegand2020drops,zhang2020mechanical,lee2020chromatin}, which can limit their growth.
	Moreover, biopolymer gels often rearrange dynamically, implying that the mechanical stress exerted by droplets can relax and they can grow further akin to the cavitation event in our model.
	Beyond our current description, the rearrangement implies visco-elastic behavior~\cite{wottawah2005optical,Broedersz2010,wen2011polymer} and biopolymer gels also often display strain-stiffening~\cite{storm2005nonlinear}.
	There is also the possibility of droplets wetting the mesh instead of excluding it completely ~\cite{ronceray2021liquid}. Taken together with the fact that the size of typical condensates is comparable to the gel's mesh size, we thus expect a rich phenomenology.
	Our theory provides a robust starting point for such future investigations.
	
\subsection*{Materials and Methods}
		The numerical simulations where performed using the \emph{py-pde} python package \cite{zwicker2020py} using an explicit Euler stepping with a second order discretization of the spatial derivative.

\subsection*{Acknowledgments}
		We thank Eric Dufresne, Pierre Ronceray, and  Robert W. Style for a critical review of the manuscript and helpful discussions. 
		For further discussions, we also thank Tal Cohen, Stefanie Heyden, and Noah Ziethen. Funding was provided by the Max Planck Society.


\printbibliography
	
	\renewcommand\thefigure{S.\arabic{figure}}   
	\renewcommand{\theequation}{S.\arabic{equation}}
	\setcounter{equation}{0}
	\setcounter{figure}{0}
	
	\newpage
	\section*{Supplementary Material}
\section{Elastic energy of a growing cavity}
We consider a droplet nucleated inside a cavity of an elastic material. This cavity has original radius $A$ and we look for the elastic energy of the system once the droplet has grown and expanded the cavity to a radius $a$.
The elastic energy $F_E$ of the system is 
\begin{equation}
F_E=\int \omega \mathrm{d}^3r\;,
\end{equation}
where $\omega$ is the energy density of the system and depends on the elastic properties of the media.
Assuming a spherical cavity and a perfectly homogenous system, the pressure $P$ exerted on the droplet is 
\begin{equation}
P = -\int\dfrac{2\sigma}{r}\mathrm{d}r\;,
\end{equation}
where $\sigma$ is the biaxial stress and $r$ the radial coordinate of the deformed solid.
We parametrize the system using the radial stretch $\lambda=r/R$, where $R$ is the radial coordinate in the original (non-deformed) solid. The two coordinates systems are related through volume conservation,
\begin{equation}
R^3 - A^3 = r^3 - a^3\;.
\end{equation}
Using the biaxial stress definition
\begin{equation}
\sigma(\lambda) = \dfrac{\lambda}{2}\dfrac{\mathrm{d}\omega}{\mathrm{d}\lambda},
\end{equation}
we obtain the pressure $P$ in terms of the energy density,
\begin{equation}
P=\int_1^{a/A}\dfrac{1}{\lambda^3 -1}\dfrac{\mathrm{d}\omega}{\mathrm{d}\lambda}\mathrm{d}\lambda\;,
\label{Eq:P_integral_form}
\end{equation}
where we have used $\lambda\rightarrow 1$ at the system's boundary.
Integrating by parts we obtain
\begin{equation}
P = \left.\dfrac{\omega}{\lambda^3 -1}\right|_1^{a/A} + \int_1^{a/A}\dfrac{3\lambda^2\omega}{(\lambda^3-1)^2}\mathrm{d}\lambda\;.
\end{equation}
Defining
\begin{equation}
P_0 = \dfrac{A^3\omega(a/A)}{a^3 - A^3},
\label{Eq:P_0def}
\end{equation}
and using the definition of Elastic Energy $F_E$, we get an expression for the elastic energy in terms of the pressure exerted on the droplet
\begin{equation}
F_E = \dfrac{4\pi(a^3 - A^3)}{3}(P - P_0).
\label{SEq:Elastic_energy}
\end{equation}
We next check whether this mechanical definition is consistent with the thermodynamic definition
\begin{equation}
\dfrac{\partial F_E}{\partial V} = P,
\label{EQ:Classical_description}
\end{equation}
where $F_E$ is differentiated with respect to the expanded cavity radius $V=4\pi a^3/3$.
Differentiating \eqref{SEq:Elastic_energy},
\begin{equation}
\dfrac{\partial F_E}{\partial V} = P - P_0 + \dfrac{(a^3 - A^3)}{3a^2}\left( \dfrac{\partial P}{\partial a} - \dfrac{\partial P_0}{\partial a}\right)\;, 
\end{equation}
using the definition of $P_0$, see \eqref{Eq:P_0def}, we find
\begin{equation}
\dfrac{\partial F_E}{\partial V} = P + \dfrac{1}{3a^2}\left((a^3 - A^2)\dfrac{\partial P}{\partial a} - A^3 \dfrac{\partial \omega}{\partial a}\right)\;.
\label{EQ:Previous_eq}
\end{equation}
Finally, using \eqref{Eq:P_integral_form}, we obtain
\begin{equation}
\dfrac{\partial P}{\partial a} = \dfrac{A^3}{a^3 - A^3}\dfrac{\partial\omega}{\partial a},
\end{equation}
which combined with \eqref{EQ:Previous_eq} recovers the thermodynamic pressure given by \eqref{EQ:Classical_description}.

\section{Free energy of a growing droplet}
We now look for a simple expression for the free energy of a growing droplet in a phase separating system. Analysing how this energy changes with volume will give us an approximation of the effect of an external pressure in this system.

Given a free energy density $f$ and a systems's size $V_\mathrm{sys}$, the total free energy is
\begin{equation}
F = Vf(\phi_\mathrm{in}) + (V_\mathrm{sys} - V)f(\phi_\mathrm{out}) + \gamma A + F_E(V)
\;,
\end{equation}
where $\phi_\mathrm{in}$ is the volume fraction inside the growing droplet, $\phi_\mathrm{out}$ is the volume fraction outside, $F_E$ is, as before, the elastic energy, $V$ is the droplet's volume, and $A$ is the droplet's surface area.

We assume material conservation in the system during the droplet's expansion, 
\begin{equation}
V_\mathrm{sys}\bar\phi = (V_\mathrm{sys} - V)\phi_\mathrm{out} + V\phi_\mathrm{in}
\;,
\end{equation}
where $\bar\phi $ is the average concentration in the system.
Assuming small changes in the dilute phase concentration, we can expand the free energy density,
\begin{equation}
f(\phi_\mathrm{out})\approx f(\phi_\mathrm{out}^0) + f'(\phi_\mathrm{out}^0)\left( \dfrac{V_\mathrm{sys}\bar\phi - V\phi_\mathrm{in}}{V_\mathrm{sys} - V} -\phi_\mathrm{out}^0\right). 
\end{equation}
We can then reorder the free energy as 
\begin{equation}
F \approx F_0 - Vg +\gamma A +F_E
\;,
\label{Eq:Approximated_FreeEnergy}
\end{equation}
with 
\begin{equation}
g = f(\phi_\mathrm{out}^0) - f(\phi_\mathrm{in}) + f'(\phi_\mathrm{out}^0)(\phi_\mathrm{in} - \phi_\mathrm{out}^0)
\end{equation}
and 
\begin{equation}
F_0 = V_\mathrm{sys} \left[f(\phi_\mathrm{out}^0) + f'(\phi_\mathrm{out}^0)(\bar\phi - \phi_\mathrm{out}^0)\right]
\;.
\end{equation}
Using the definition of osmotic pressure,
\begin{equation}
\Pi = -f(\phi) + f'(\phi)\phi
\;,
\end{equation}
we express $g$ using the difference in osmotic pressures, 
\begin{equation}
g = \Pi_\mathrm{in} - \Pi_\mathrm{out}^0 - \left[f'(\phi_\mathrm{in}) -     f'(\phi_\mathrm{out}^0)\right]\phi_\mathrm{in}
\;.
\end{equation}
Finally, using the definition of chemical potential $\mu = \nu f'(\phi)$, with $\nu$ the molecular volume, we obtain
\begin{equation}
g = \Pi_\mathrm{in} - \Pi_\mathrm{out}^0 - \left[\mu_\mathrm{in} - \mu_\mathrm{out}^0\right]c_\mathrm{in}
\;,
\label{Eq:driving_strength}
\end{equation}
where $c_\mathrm{in}=\phi_\mathrm{in}/\nu$ is the number concentration inside the droplet. In the case of an incompressible dense phase, i.e. where $\phi_\mathrm{in}$ is constant, we find that the driving strength $g$ is independent of droplet volume.

\section{Stability analysis}
We showed in the previous section that the driving strength~$g$ is typically independent of  the droplet volume~$V$.
Phase separation is favorable (in the absence of surface tension and elastic effects), when $g>0$. 
To see how elastic effects affect the phase separation, we differentiate \eqref{Eq:Approximated_FreeEnergy} with respect to the droplet volume,
\begin{equation}
\dfrac{\partial F}{\partial V} = -g + \dfrac{2\gamma}{R} + P_E
\;,
\end{equation}
where we have used \eqref{EQ:Classical_description} to derive the elastic pressure.
Therefore, droplet growth is favourable if
\begin{equation}
g> \dfrac{2\gamma}{R} + P_E
\;.
\end{equation}
A droplet will thus grow as long as the driving strength $g$ is bigger than the total pressure difference between the inside and outside of the droplet.

We next study the stability of a steady state, which exists when $g = P(R^*)$.
This state is stable if
\begin{equation}
\left.\dfrac{\partial^2 F}{\partial V^2}\right|_{R=R^*} =\left. \dfrac{\partial}{\partial V}\left( \dfrac{2\gamma}{R}+ P_E(R)\right)\right|_{R=R^*}  >0
\;.
\label{Eq:Stability_condition}
\end{equation}
Therefore, a droplet will be stable if the pressure increases with increasing radius. 

The same stability condition can be obtained from the dynamical equations presented in the main manuscript,
\begin{equation}
\frac{\mathrm{d}R_i}{\mathrm{d}t} =
\frac{D}{R_i c_\mathrm{in}}\left[
c(\vec{x}_i)-c_\mathrm{eq}\bigl(P(R_i), T\bigr)
\right]
\;.
\end{equation}
A linear stability analysis shows that the droplet radius is stable if $c_\mathrm{eq}$ increases with droplet radius. Since $c_\mathrm{eq}$ is a monotonically increasing function of $P(R)$, the stability condition reduces to
\begin{equation}
\left. \dfrac{\partial}{\partial R}\left( \dfrac{2\gamma}{R}+ P_E(R)\right)\right|_{R=R^*}  >0
\;,
\end{equation}
which is equivalent to Eq. \eqref{Eq:Stability_condition}.

\section{Droplet nucleation}
We here study the droplet nucleation behavior by considering the energy necessary to cross the nucleation barrier in the absence of elastic effects.
We consider a supersaturated homogeneous solution with concentration $c_0$ and use the driving strength $g$, Eq. \ref{Eq:driving_strength}, to estimate the energy change due to a droplet nucleating.
We assume that the differences in pressure relax much more quickly than the differences in chemical potential, implying that we can approximate the driving strength as
\begin{align}
g\approx [\mu_\textrm{out} - \mu_\textrm{in}]\cIn
\;.
\end{align}
To estimate its value, we thus need to determine the chemical potentials~$\mu_\mathrm{out}$ and $\mu_\mathrm{in}$  outside and inside the droplet, respectively.

Assuming that the nucleated droplet is small, the chemical potential outside is close to that of the homogeneous phase, which can be estimated using ideal solution theory, $\mu_\mathrm{out}\approx \kBT\log c_\mathrm{out}$.
To obtain an upper bound on the nucleation rate, we seek the strongest possible driving strength and thus use $c_\mathrm{out} \approx c_0$.

We estimate the chemical potential~$\mu_\mathrm{in}$ inside the droplet by considering the phase separated system, which reached an equilibrium between a dilute phase with concentration $c_\mathrm{out}^\mathrm{eq}$ and a droplet phase with concentration $c_\mathrm{in}^\mathrm{eq}$.
In equilibrium, the chemical potential of both phases are identical, and we thus have
\begin{equation}
\mu_\mathrm{in}^\mathrm{eq} = \mu_\mathrm{out}^\mathrm{eq} \approx \kBT \log(c_0-\Delta c  ),
\end{equation}
where $c_0-\Delta c$ is a lower bound for the concentration in the dilute phase.

Finally, we assume that the chemical potential inside the droplet is constant during equilibration, $ \mu_\mathrm{in} \approx  \mu_\mathrm{in}^\mathrm{eq}$, to obtain the estimate
\begin{equation}
g \approx \kBT\cIn\log{\left(\dfrac{c_0}{c_0-\Delta c}\right)}
\;.
\end{equation}
Consequently, the change in free energy $\Delta\mathcal{F}$ due to one droplet of radius $R$ is 
\begin{equation}
\Delta\mathcal{F}\approx 4\pi R^2\gamma -\dfrac{4\pi}{3}R^3g
\;,
\end{equation}
where $\gamma = \unit[4.4]{mN/m}$ is the surface tension measured in the experiments~\cite{style2018liquid}.
The location of the maximum of this curve corresponds to the nucleation radius
\begin{equation}
R_\textrm{nuc} = \dfrac{2\gamma}{g}\approx \unit[3.09]{nm}
\;,
\end{equation}
where we have used the measured value $\cIn\kBT=\unit[11]{MPa}$~\cite{rosowski2020elastic}.
The associated energy barrier is
\begin{equation}
\Delta\mathcal{F}_\mathrm{nuc}=\dfrac{16\pi\gamma^3}{3g^2} \approx \unit[1.76\cdot10^{-19}]{J}\approx 42.4\hspace{2 pt}\kBT
\;.
\end{equation}
The probability $P_\textrm{nuc}$ of droplet nucleation can then be estimated using classical nucleation theory,
\begin{equation}
P_\textrm{nuc} = k e^{-\Delta\mathcal{F}_\mathrm{nuc}/\kBT}
\approx k \cdot 10^{-19}
\;.
\end{equation}
The pre-factor~$k$ in this theory is difficult to estimate, but this expression shows that the nucleation rate is suppressed by $10^{-19}$ and homogeneous nucleation is thus very unlikely in this system. 

The fact that droplets appear in the experiments suggests that they are nucleated by alternative paths.
We thus propose that heterogeneous nucleation is crucial.
In heterogeneous nucleation droplets are nucleated around surfaces or imperfections, which effectively lower the nucleation barrier.
The interaction of the droplet with these nucleation sites scales with the contact area, which scales with $\propto R^2$ if the size of the nucleation site is about $R_\mathrm{nuc}$ or larger.
Consequently, the primary effect of heterogeneous nucleation is to lower the effective surface tension.
We thus assume that our system for small droplets has an effective surface tension that is smaller than the measured surface tension for large cavitated droplets.
We thus do not discuss surface tension effects in the main text and rather assume that droplet nucleation happens quickly.

\section{Results are independent of mesh size and cavitation radius}

\begin{figure}[htb]
	\centering
	\includegraphics[width=.8\columnwidth]{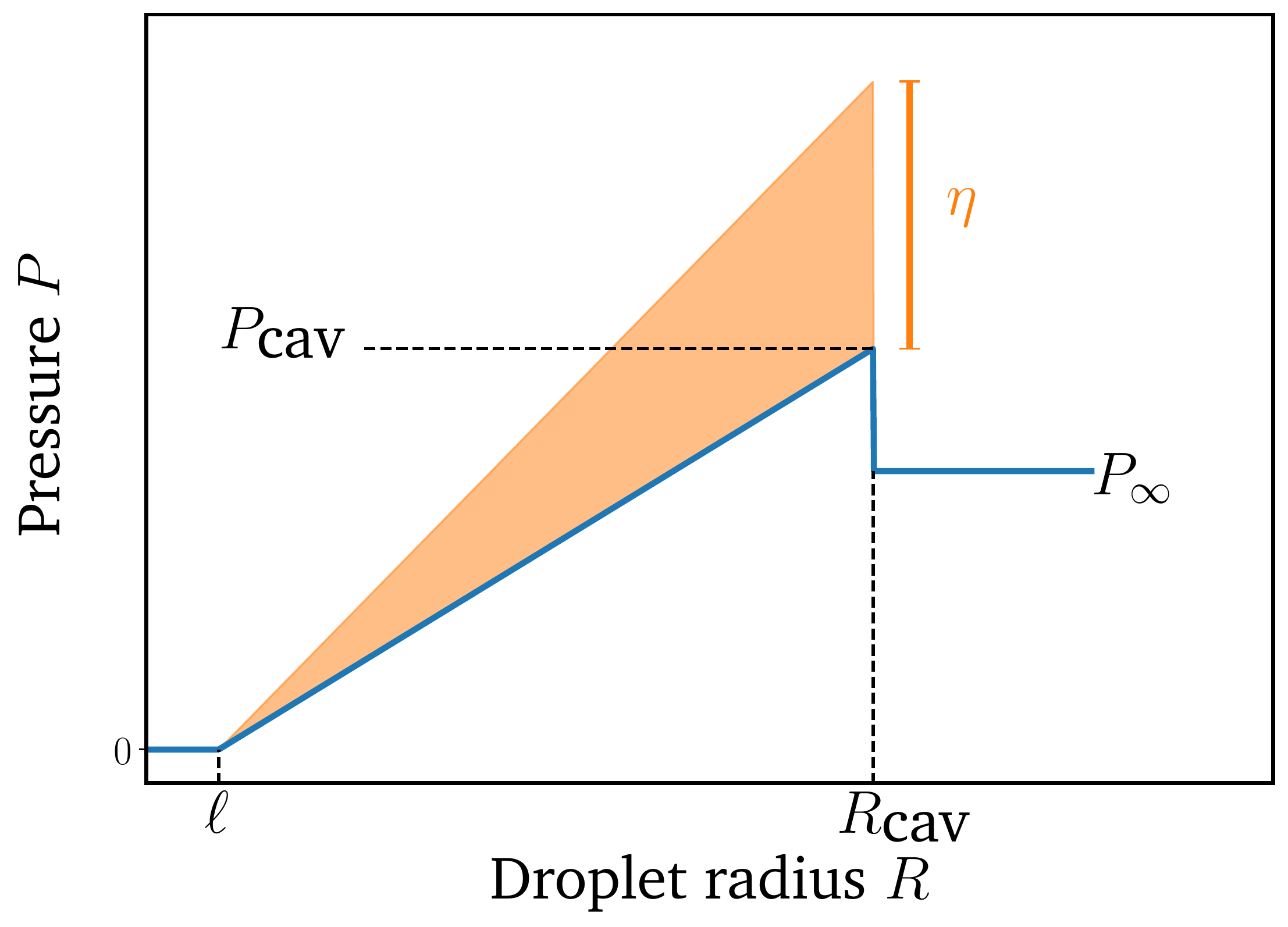}
	\caption{\textbf{Stress-strain relation of the Breakage model.} 
		Pressure~$P$ as a function of the droplet radius~$R$, which is imposed in numerical simulations.
		The plot indicates the mesh size $\ell$, the radius $R_\textrm{cav}$ at which the droplet cavitates, the maximal pressure $P_\textrm{cav}$ the mesh can exert, and the pressure $P_\infty$ after breakage. The shaded area shows the possible values for $P_\textrm{cav}$ given the distribution parameter $\eta$.}
	\label{Fig:Supp_pressure_curve}
\end{figure}

\begin{figure}[htb]
	\centering
	\includegraphics{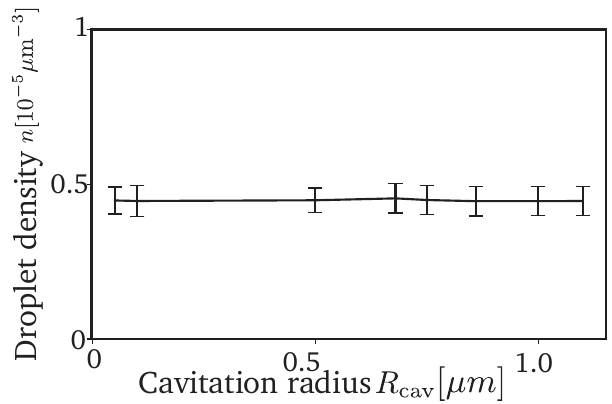}
	\caption{\textbf{Droplet density is independent of cavitation radius} Density $n$ of cavitated droplets as a function of the cavitation radius $R_\mathrm{cav}$. Model parameters as in Fig. 2 of the main manuscript, except $\eta/m$ = \unit[3$\cdot$10\textsuperscript{5}$E$]{\textmu m\textsuperscript{3}}.
	}
	\label{Fig:Cavitation_dependency}
\end{figure}

In the Breakage model, we use a simplified pressure curve, which is parameterized by the mesh size~$\ell$, the cavitation radius~$R_\mathrm{cav}$, the cavitation pressure~$P_\mathrm{cav}$, the heterogeneity parameter~$\eta$, and the final pressure~$P_\infty$ for large radii; see \figref{Fig:Supp_pressure_curve}.
The mesh size $\ell$ only determines when the pressure starts to increase, and therefore the slope of the pressure curve.
Considering the cavitated droplet density $n$, introducing variation in the mesh size $\ell$ is equivalent to variations in  $P_\textrm{cav}$ for different droplets.
Beside this, varying the mesh size $\ell$ changes the size of the small droplets, but since they do not affect the number of cavitated droplets, we do not study this effect further.
We thus for simplicity only vary $P_\textrm{cav}$ and keep $\ell$ the same for all droplets.
Moreover, numerical simulations indicate that the cavitation radius $R_\textrm{cav}$ has basically no influence on the cavitated droplet density $n$; see Fig. \ref{Fig:Cavitation_dependency}. 
Given these results we choose not to vary $R_\textrm{cav}$ between droplets or simulations, and keep $\eta/m$ as our only free parameter to fit experimental data.

\section{Ostwald Ripening}
In this section we estimate the relevant timescale for Ostwald ripening in our system to asses its relevance for the cavitated droplets.
Ostwald ripening is driven by surface tension and the relevant scale is the capillary length scale $\ell_\gamma$ of the system \cite{weber2019physics},
\begin{equation}
\ell_\gamma = \dfrac{2\gamma}{\cIn \kBT}
\approx \unit[0.8]{nm}
\;,
\end{equation}
where we have used the surface tension $\gamma = \unit[4.4]{mN/m}$ of macroscopic droplets~\cite{style2018liquid}. Linear stability analysis shows that the fastest growing mode $\lambda$ is given by~\cite{Zwicker2015}
\begin{equation}
\lambda = \dfrac{D\ell_\gamma}{R^3}\dfrac{\cEq}{\cIn}
\;.
\end{equation}
Considering a typical radius $R$ = \unit[10]{\textmu m}  of a cavitated droplet, together with $D$ = \unit[50]{\textmu m\textsuperscript{2}/s} and $\cEq/\cIn\approx  c_\textrm{sat}(\unit[300]{K})/\cIn= 0.054$, we find $\lambda\approx$\unit[2.14$\cdot$ 10\textsuperscript{-6}]{s\textsuperscript{-1}}.
Consequently, the timescale $\tau = \lambda^{-1}$ of Ostwald ripening between the cavitated droplets is 
\begin{equation}
\tau \approx \unit[130]{hr}
\;.
\end{equation}
Interestingly, the same expression is obtained when considering the critical radius $R_c$ of the Lifshitz–Slyozov scaling law \cite{weber2019physics}
\begin{equation}
R_c \propto \left(\dfrac{D\ell_\gamma\cEq}{\cIn} t \right)^{1/3}
\;. 
\end{equation}
We thus conclude that Ostwald ripening is a slow process in our system and we can neglect it. 

\section{Numerical Simulations with Nucleation}

\begin{figure}[tb]
	\centering
	\includegraphics[width=\columnwidth]{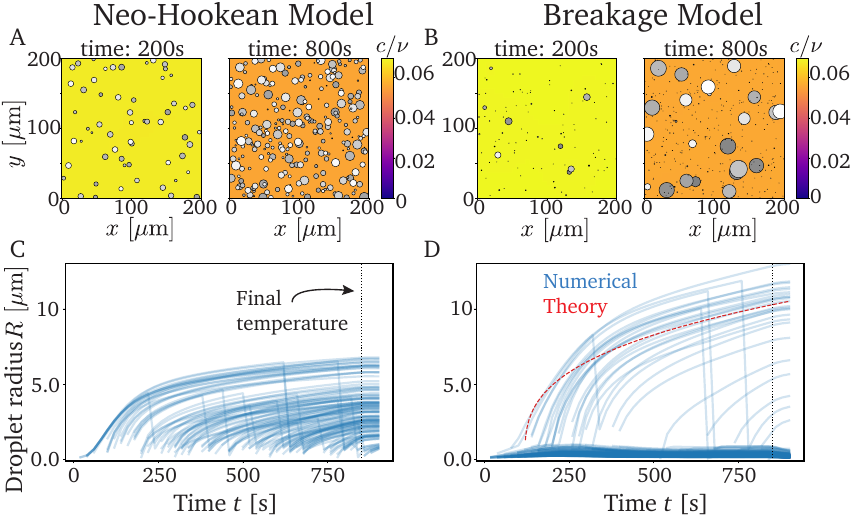}
	\caption{\textbf{Breakage stops newly nucleated droplets from growing.} (A,B) Snapshots of a typical simulation with a constant nucleation rate for Neo-Hookean (left) and Breakage pressure curves (right). Disks indicate droplets, while the heat map indicates the concentration~$c$ in the dilute phase, gray scale indicates depth in the z-axis. (C,D) Droplet radii over time for all nucleated droplets. $p_\mathrm{nuc}$ = \unit[0.5]{s\textsuperscript{-1}}, other parameters as in Figure 2 of the main text.}
	\label{Fig:Supp_nucleation}
\end{figure}

To make a more realistic comparison of the Neo-Hookean model and the Breakage model, we performed numerical simulations with a  simple droplet nucleation protocol. The simulations were initialized without any droplets and at each time step a droplet might nucleate with a probability $p_\mathrm{nuc}$. The new droplet appears with $R<\ell$.
The typical simulations presented in Figure \ref{Fig:Supp_nucleation} show that in the Neo-Hookean model the new droplets can grow freely, leading to a wide range of final radii.
In contrast, in the Breakage model, most new droplets get stuck at mesh size and do not grow further, thus producing a monodispersed emulsion of cavitated droplets.  

\section{Pair correlation function}
To quantify the position correlations of droplets in our system we defined the pair correlation function $g(r)$ which gives the probability of finding a droplet at distance $r$ from the reference droplet. It is formally defined as
\begin{equation}
g(r)= \dfrac{V_\mathrm{sys}}{N}\left\langle\sum_i \delta(\vec{r}-\vec{r_i})\right\rangle\;,
\end{equation}
where the sum is over all cavitated droplets, the average is over different ensembles, $N$ is the number of cavitated droplets, and $V_\mathrm{sys}$ the system's volume. In practice we calculate it as
\begin{equation}
g(r)= \dfrac{h(r)V_\mathrm{sys}}{4\pi N^2 r^2 \Delta r}\;,
\end{equation}
where $h(r)$ is the histogram of the distances between droplets (a total of $N(N-1)$ elements) and $\Delta r$ is the histogram's bin size. Finally, to collapse the different curves we normalized by the mean droplet distance $2[3/(4\pi n)]^{1/3}$.

\begin{figure}[tb]
	\centering
	\includegraphics[width=\columnwidth]{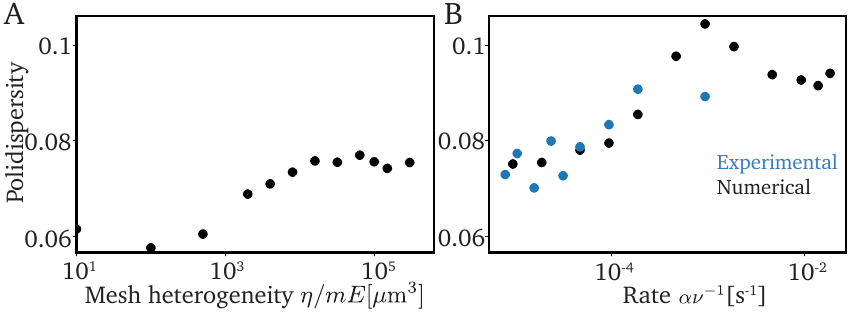}
	\caption{\textbf{Polydispersity increases at higher droplet densities.} Polydispersity, defined as the standard deviation of the droplet radius divded by its mean, for (A) different mesh heterogeneities $\eta/mE$ and (B) different rates $\alpha$. Experimental data from  \cite{style2018liquid}.  }
	\label{Fig:Polydispersity}
\end{figure}

\begin{figure*}[tb]
	\centering
	\includegraphics[width=\textwidth]{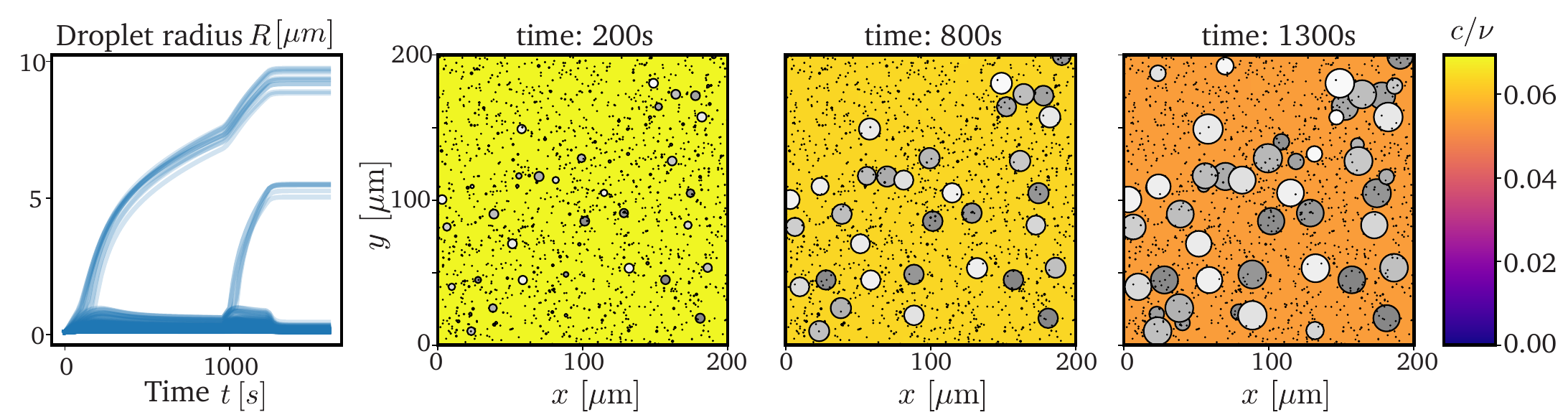}
	\caption{\textbf{Double quench experiment produces a bimodal distribution.} Left panel: Droplet radii as a function of time displaying a bimodal distribution of the cavitated droplets and several small ones kept at mesh size. Subsequent panels: 2-D projection of a typical time evolution in this system, showing droplets starting to cavitate, grown first group of cavitated droplets, and final state of the system. Smaller droplets are shown as black dots for didactic purposes and might appear as inside cavitated droplets due to 3-D projection. Parameters are $\eta/m =$ \unit[3$\cdot$10\textsuperscript{5}$E$]{ \textmu m \textsuperscript{3}}, $E =$ \unit[80]{kPa}, first cooling rate $\alpha =$ \unit[7.77$\cdot$10\textsuperscript{-6}]{s\textsuperscript{-1}$\nu$\textsuperscript{-1}}, second cooling rate $\alpha =$ \unit[3.11$\cdot$10\textsuperscript{-5}]{s\textsuperscript{-1}$\nu$\textsuperscript{-1}}. Other parameters as in Figure 2. }
	\label{Fig:Double_quench}
\end{figure*}

\section{Theoretical estimation of cavitated droplet density $n$}
We here provide details on the analytical theory to predict the density~$n$ of cavitated droplets.
This is based on the simplified picture that a single large droplet depletes a surrounding volume of radius~$L$ in a diffusion limited process.
To obtain the concentration field around this droplet, we solve Eq. (6) in a spherical domain of radius $L=\left[3/(4\pi n)\right]^{1/3}$ with boundary conditions $\partial_rc|_{r=L}=0$ and  $c(r=R_\textrm{cav})=c_\textrm{sat}(t)\exp{\left(P_\infty/(c_\textrm{in}k_BT)\right)}$.
Defining $\bar{c} = c + \alpha t\exp{\left(P_\infty/(c_\textrm{in}k_BT)\right)}$ turns Eq. (6) into
\begin{equation}
\partial_t \bar{c} = D \nabla^2 \bar{c} + \alpha\exp{\left(\dfrac{P_\infty}{c_\textrm{in}k_BT}\right)}  - c_\mathrm{in}\sum_i \dfrac{\mathrm{d}V_i}{\mathrm{d}t} \delta(\vec{x_i} - \vec{x})
\;,
\label{Eq:phi_dynamics_withSource}
\end{equation}
with the simpler boundary conditions $\partial_r\bar{c}|_{r=L}=0$ and  $\bar{c}(r=R_\textrm{cav})=c_0\exp{\left(P_\infty/(c_\textrm{in}k_BT)\right)}$. Assuming radial symmetry and steady state, the field around a cavitated droplet of radius $R_\textrm{cav}$ reads x`
\begin{equation}
\begin{split}
\bar{c}(r) =& \left[c_0 + \dfrac{\alpha}{6D}\left(R_\textrm{cav}^2 - r^2\right) \right.\\
&+\left. \dfrac{\alpha L^3}{3D}\left(\dfrac{1}{R_\textrm{cav}}-\dfrac{1}{r}\right)\right]\exp{\left(\dfrac{P_\infty}{c_\textrm{in}k_BT}\right)}
\;.
\end{split}
\label{Eq:phi_of_r}
\end{equation}
We thus simplified the cavitation scenario by assuming that once a droplet cavitates it absorbs material in a sphere of radius $L$ around them. 

To account for heterogeneity in the cavitation pressures~$P_\textrm{cav}$, we describe them through their cumulative distribution function $F(P_\textrm{cav})$, which gives the fraction of droplets whose cavitation pressure is lower than $P_\textrm{cav}$. Since only droplets with the lowest cavitation thresholds will cavitate, we only need to describe the lower end of $F(P_\textrm{cav})$. Since there must be a lowest, positive cavitation pressure~$P_\mathrm{cav}^\mathrm{min}$, we assume a linear expansion around this minimum,
\begin{equation}
\mathcal{F}(P_\textrm{cav})=\dfrac{P_\textrm{cav} - P_\mathrm{cav}^\mathrm{min}}{\eta}\Theta(P_\textrm{cav} - P_\mathrm{cav}^\mathrm{min})\;,
\label{Eq:Cummulative_distribution_SI}
\end{equation}
where $\eta$ describes how widely distributed the lower thresholds are. This expression can be rewritten using Eq.(5) to obtain the cumulative distribution in terms of the equilibrium concentrations.
\begin{equation}
\begin{split}
\mathcal{F}(c,t) = &\left[\dfrac{c_\mathrm{in}k_BT}{\eta}\log{\left(\dfrac{c}{c_0-\alpha t }\right)} - \dfrac{P_\mathrm{cav}^\mathrm{min}}{\eta}\right]\\
&\Theta\left( c_\mathrm{in}k_BT\log{\left(\dfrac{c}{c_0-\alpha t}\right)}-P_\mathrm{cav}^\mathrm{min}\right)
\;.
\end{split}
\end{equation}
Therefore, given a density $m$ of nucleated droplets, the density of droplets with cavitation threshold below $c$ is $m\mathcal{F}(c,t)$.
For this theory to be self-consistent, the  aforementioned volume $V=n^{-1}$ must have only one droplet with a cavitation threshold below the concentration field, i.e.
\begin{equation}
1=4\pi\int_0^L m \mathcal{F}(c(r),\bar t)r^2 dr.
\end{equation}
Here, we take the time $\bar{t}$ such that the equilibrium concentration of the $n$-th droplet matches the total amount of material in the system $c_0$,
\begin{equation}
\bar{t} = \dfrac{c_0}{\alpha}\left[1 -  \exp{\left(\dfrac{-P_\mathrm{cav}^\mathrm{min}-\eta n /m }{c_\textrm{in}k_BT}\right)}\right]\;.
\label{Eq:bar_t}
\end{equation}
We can now combine \eqref{Eq:phi_of_r}-\eqref{Eq:bar_t} to obtain an implicit relation for the cavitated droplet density $n$, which we solve numerically to obtain the lines shown in the main text.

\begin{figure}[tb]
	\centering
	\includegraphics[width=\columnwidth]{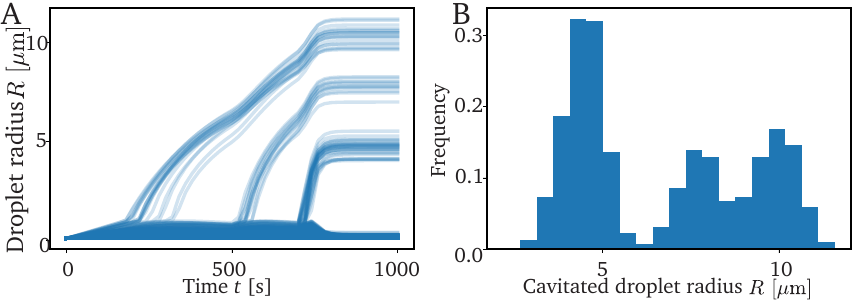}
	\caption{\textbf{Increasing rate stepwise leads to increasingly more droplet cavitating.} Simulation increasing the material rate $\alpha$ twice during the cooling process. A) Droplet radii as a function of time displaying a trimodal distribution of the cavitated droplets and several small ones kept at mesh size. B) Radii probability distribution. Parameters are $\eta/m =$ \unit[3$\cdot$10\textsuperscript{5}$E$]{ \textmu m \textsuperscript{3}}, first cooling rate $\alpha =\alpha_0=$ \unit[7.77$\cdot$10\textsuperscript{-6}]{s\textsuperscript{-1}$\nu$\textsuperscript{-1}}, second cooling rate $\alpha =4\alpha_0$, and third cooling rate $\alpha= 16\alpha_0$. Other parameters as in Figure 2. }
	\label{Fig:Triple_quench}
\end{figure}

\section{Temperature protocol changes}
Numerical simulations, where the cooling rate is increased after droplet cavitation, show that a new group of droplets can cavitate if the new rate is high enough. Examples of this type of simulations are shown in \figref{Fig:Double_quench} and \figref{Fig:Triple_quench}. We show in \eqref{Eq:phi_of_r} that the concentration profile around a cavitated droplet depends strongly on the material rate $\alpha$. If this rate is suddenly increased, the new concentration profile might be higher than the cavitation concentration $c_\textrm{cav}$ of some of the small droplets, thus causing their cavitation. This process can then be repeated to cavitate additional droplets, as depicted in \figref{Fig:Triple_quench}. Therefore, controlling the rate~$\alpha$ allows to control the droplet size distribution.

\end{document}